\documentclass[10pt,conference]{IEEEtran}
\IEEEoverridecommandlockouts
\usepackage{url}
\usepackage{multirow}
\usepackage{makecell}
\usepackage{booktabs}
\usepackage{cite}
\usepackage{amsmath,amssymb,amsfonts}
\usepackage{algorithmic}
\usepackage{graphicx}
\usepackage{textcomp}
\usepackage{xcolor}
\def\BibTeX{{\rm B\kern-.05em{\sc i\kern-.025em b}\kern-.08em
    T\kern-.1667em\lower.7ex\hbox{E}\kern-.125emX}}

\usepackage{csquotes}
\usepackage{tcolorbox}
\usepackage{sparklistings}  
\tcbuselibrary{listings, breakable}
\usepackage{xcolor}

\newtcblisting{sparksbox}[1]{
  listing only,
  listing remove caption=false,
  colback=gray!3,
  colframe=gray!60,     
  boxrule=0.4pt,        
  arc=1pt,              
  sharp corners,
  listing options={
    language=sparks,
    style=change,
    basicstyle=\scriptsize\ttfamily,
    columns=fullflexible,
    keepspaces=true,
    showstringspaces=false,
    #1
  }
}

\newcommand{\TB}{\mathop{\mathrm{TB}}}

\newcommand{\PP}{{\textsc{P}}}
\newcommand{\Sparks}{\texttt{Sparks}}
\newcommand{\Sparktope}{\textsc{sparktope}}
\newcommand{\Asm}{\texttt{Asm}}
\newcommand{\gurobi}{\textsc{gurobi}}

\newcommand{\cplex}{\textsc{cplex}}
\newcommand{\ibmilog}{\textsc{ibm ilog}}
\newcommand{\scip}{\textsc{scip}}

\newcommand{\BONE}{\mathit{b1}}
\newcommand{\BTWO}{\mathit{b2}}
\newcommand{\BTHREE}{\mathit{b3}}
\newcommand{\iEight}{\mathit{I8}}
\newcommand{\fOne}{\mathit{F1}}
\newcommand{\fTwo}{\mathit{F2}}
\newcommand{\vLA}{\mathit{a}}
\newcommand{\tbLocal}{\operatorname{TB}}
\newcommand{\distParent}{\operatorname{dist\!Parent}}

\newcommand{\parent}{\operatorname{parent}}
\newcommand{\children}{\operatorname{children}}
\newcommand{\prevSiblings}{\operatorname{prev\!Sib}}

\newcommand{\ETIstart}{\operatorname{ETI}_{\mathsf{start}}}
\newcommand{\ETIend}{\operatorname{ETI}_{\mathsf{end}}}
\newcommand{\ETI}{\operatorname{ETI}}
\newcommand{\ETIG}{\operatorname{ETIG}}
\newcommand{\nonRootAncestors}{\operatorname{nr\!Anc}}
\newcommand{\distPenultimateToSA}{\operatorname{distPenul}}
\newcommand{\distCumulativeToTarget}{\operatorname{distCumul}}
\newcommand{\vMi}{\mathit{max\!Iter}} 
\newcommand{\vMaxIterTuple}{\mathbf{Max\!Iter}} 

\newif\ifanonymous
\anonymousfalse 
\begin{document}
\title{Efficient Compilation of Algorithms into Compact Linear Programs }

\ifanonymous
\else
\author{\IEEEauthorblockN{Shermin Khosravi}
\IEEEauthorblockA{\textit{Faculty of Computer Science} \\
\textit{University of New Brunswick}\\
Fredericton, Canada \\
shermin.khosravi@unb.ca}
\and
\IEEEauthorblockN{David Bremner}
\IEEEauthorblockA{\textit{Faculty of Computer Science} \\
\textit{University of New Brunswick}\\
Fredericton, Canada \\
bremner@unb.ca}
}
\fi

\maketitle

\begin{abstract}
While data-driven approaches such as Machine Learning and Artificial Intelligence continue to find new applications across many domains, traditional mathematical optimization frameworks remain highly effective for solving real-world problems having well-defined constraints. 
In particular, Linear Programming (LP) is widely applied in industry and is a key component of various other mathematical problem-solving techniques. 
Consequently, LP has attracted significant research interest, particularly in exploring its expressive powers as a computational tool. 
Recent work introduced an LP compiler translating polynomial-time, polynomial-space algorithms into polynomial-size LPs using intuitive high-level programming languages, offering a promising alternative to manually specifying each set of constraints through Algebraic Modeling Languages (AMLs). However, the resulting LPs, while polynomial in size, are often extremely large, posing challenges for existing LP solvers.
In this paper, we propose a novel approach for generating substantially smaller LPs from algorithms. 
Our goal is to establish minimum-size compact LP formulations for problems in $\PP$ having natural formulations with exponential extension complexities.
Our broader vision is to enable the systematic generation of Compact Integer Programming (CIP) formulations for problems with exponential-size IPs having polynomial-time separation oracles.
To this end, we introduce a hierarchical linear pipelining technique that decomposes nested program structures into synchronized regions with well-defined execution transitions---functions of compile-time parameters. This decomposition allows us to localize LP constraints and variables within each region, significantly reducing LP size without the loss of generality, ensuring the resulting LP remains valid for all inputs of size~$n$.
We demonstrate the effectiveness of our method on two benchmark problems---the makespan problem, which has exponential extension complexity, and the weighted minimum spanning tree problem---both of which have exponential-size natural LPs. Our results show up to a $25$-fold reduction in LP size and substantial improvements in solver performance across both commercial and non-commercial LP solvers.
\end{abstract}
\section{Introduction}
\label{sec:introduction}
Although data-driven methods, such as Machine Learning and Artificial Intelligence, continue to find new applications, traditional optimization methods remain highly effective for certain real-world problems, particularly when problem-specific constraints are either known or can be extracted from data. For instance, the winning solution of Amazon's Last Mile Research Challenge, which significantly outperformed alternative solutions, was based on traditional optimization techniques rooted in operations research~\cite{cook_constrained_2024}.

Among these, Linear Programming (LP) is an optimization technique for solving problems defined by linear constraints and objectives. It has a wide range of applications across diverse fields and is at the core of other optimization techniques such as Integer Programming (IP). LP's pivotal role and widespread applicability have led to the recognition of the simplex algorithm---a well-known LP solving method---as one of the top ten algorithms of the $20$th century~\cite{dongarra_guest_2000}. 

In particular, an active area of research focuses on determining the minimum LP size required to accurately represent $\PP$ problems lacking known natural\footnote{Formulations based on the problem description, such as a polytope defined by the convex hull of characteristic vectors of a decision problem with a \enquote{yes} answer} polynomial size LPs. 
Although polynomial-size LPs have been discovered for some of these problems, there remains interest in finding even smaller LP formulations~\cite{williams_linear-size_2002, fiorini_smaller_2017, aprile_smaller_2021, abdelmaguid_efficient_2018}). These efforts not only improve computational efficiency but also achieve tighter relaxations for IP problems and may provide deeper theoretical insights.

While exponential-size LPs having polynomial-time separation oracles can be solved in polynomial time using the Ellipsoid method~\cite{grotschel_ellipsoid_1981}, the method is rarely used in practice due to its slow convergence. Consequently, finding equivalent but smaller LP formulations eliminates the need for execution of separation algorithms and allows the use of more practical LP algorithms that require a full description of the model~\cite{carr_compact_2002}.
Unlike with LPs, where larger formulations generally increase solver difficulty~\cite{ted_integer_2025}, some large IPs can be solved relatively quickly, while certain small IPs remain unsolved to this day. Nevertheless, since IP solvers mainly rely on repeatedly solving LP relaxations, the model size affects the computational cost of each iteration. This effect is especially pronounced in algorithms that solve the LP relaxation from scratch at every iteration~\cite{karwan_redundancy_1983}. In branch-and-cut methods~\cite{zhang_survey_2023}, runtime increases gradually as additional constraints are generated lazily~\cite{talbi_parallel_2006}. If multiple cuts are added per iteration~\cite{fabricio_oliveira_optimisation_2025}, reusing dual information from previous solves can become nearly as costly as solving the LP from scratch.
Methods that rely on separation oracles\footnote{A black box that, given a point in the solution space, determines whether the point lies within a certain convex set and, if not, returns a hyperplane that separates the point from the set} to iteratively narrow down the search often operate on a presolved model. As a result, certain presolving techniques that prevent the translation between presolved and original variables must be disabled to preserve compatibility with the oracle interface~\cite{ibm_presolve_2021, ibm_reformulations_2021}.
Moreover, for certain problems---such as the contact map overlap problem~\cite{carr_compact_2002}---the cuts generated by the separation oracle at each iteration yield only small improvements in the objective value, leading to long runtime due to the large number of iterations.

Although Algebraic Modeling Languages (AMLs) help with expressing optimization models in a more abstract way, users still have to manually define each set of LP constraints from scratch, which is a tedious task, particularly for more complex models. 
A recent study~\cite{avis_sparktope_2022} introduces an LP compiler that converts algorithms into LP formulations. When the input algorithm runs in polynomial time and space, the resulting LP is guaranteed to be of polynomial size. 
This approach allows users to model problems using more intuitive high-level programming languages and enables the construction of polynomial-size LPs for problems in $\PP$ whose natural polytopes have exponential extension complexity.
However, the resulting LPs are often exceptionally large---even for relatively small algorithms---posing challenges for solvers due to memory limitations, numerical stability, and rounding errors.
A significant contributor to the LP's size growth is how the compiler handles the black-box nature of program execution at compile time. 
To account for all $2^n$ possible execution paths for inputs of size $n$, the compiler unrolls each line of code across $\TB(n)$ time steps, where $\TB(n)$ is a user-provided upper bound on the number of execution steps. This process generates $\TB(n)$ single-step execution paths per code line, allowing control flow transitions at any time step. 
Consequently, the model uniformly assumes potential $\TB(n)$ execution frequency for all code lines, even though many of these execution paths may be infeasible or unreachable at runtime.

In this paper, we address the black-box nature of execution patterns at compile time by proposing a method inspired by static analysis in abstract interpretation~\cite{cousot_abstract_2014}. Just as abstract interpretation infers reachability using interval abstractions, our approach computes conservative \emph{execution time intervals} (ETIs), specifying the time ranges during which each code line can be reachable. 
To enforce these intervals, we draw on the concept of synchronization barriers from multithreaded programming, where execution pauses until all threads reach a designated barrier. 
Our method automatically decomposes the program---guided by syntactic structure and type information---into a hierarchy of linear pipelines. These pipelines are synchronized with a global clock using the statically known upper bounds on the execution time of each block. 
The barriers impose temporal ordering based on hierarchical dependencies, introducing structural invariants over the execution time. As a result, execution traces are determined not only by program logic and input, but also by barrier-imposed time intervals---which include idle periods for blocks that are inactive during certain time steps.
Although execution proceeds as a single, sequential pass without parallelism or concurrency, the pipeline eliminates redundant execution paths by generating a single set of transition constraints between blocks.
This allows the compiler to generate LP constraints and variables only for the ETIs of blocks that contain the corresponding code statements and variables. In addition to reducing LP size, this method also introduces variation in execution frequencies at the block level within the LP model---an effect that can be further exploited by compiler optimization techniques sensitive to execution frequency, some of which are discussed in Section~\ref{sec:relatedWork}. 
We refer to our approach as Hierarchical Synchronization Barriers (HSB).
To efficiently compute ETIs, HSB constructs a tree that captures the hierarchical structure of the code, annotating each block with time-related metadata in the form of constants and symbolic formulas. 
This information is then used to generate ETIs---and their unions over selected subsets---on demand during LP constraint and variable generation, enabling the elimination of some redundancies in the LP.
Although synchronization introduces additional compile-time parameters---previously specified manually by the user when calculating $\TB(n)$ manually in the original compiler---it enables the compiler to compute $\TB(n)$ automatically, requiring the user to provide only the maximum number of loop iterations as parameters. 
Consequently, unlike in abstract interpretation, HSB does not require termination abstractions such as widening~\cite{cortesi_widening_2011}, since iteration bounds are known at compile time.
In effect, HSB transforms the general-purpose imperative source language \Sparks{} into an embedded Domain Specific Language (eDSL) designed for generating efficient LP formulations and for semi-automatically computing the problem's $\TB(n)$.
It achieves this by inserting annotations---either automatically or via user specification---into the source code. 
In the automatic case, the user still interacts with the eDSL indirectly by providing a parameter file that specifies upper bounds on problem size, just as in the original compiler. 
These annotations alter the program's runtime trace pattern in the IR without altering its semantics, optimizing it specifically for LP generation. 
These modifications structure execution into a sequence of smaller, well-defined black boxes with statically known transition times. As a result, each line of code becomes reachable only within the ETIs of its enclosing block. 
Our LP size reduction techniques rely on both the knowledge of the program structure and the design of the LP constraints. As such, they are not achievable through presolving techniques or conventional compiler optimizations in isolation.

The remainder of the paper is organized as follows. Section~\ref{sec:background} provides an overview of the LP compiler \Sparktope{}~\cite{avis_sparktope_2022}. Section~\ref{sec:relatedWork} surveys existing methods for reducing LP size and highlights their key limitations, particularly those that efficient algorithm-to-LP conversions seek to overcome. Section~\ref{sec:methodology} introduces our proposed HSB methods. Section~\ref{sec:results} presents experimental results on two benchmark problems. Finally, Section~\ref{sec:conclusion} concludes the paper and outlines future directions of this work.

\section{Background}
\label{sec:background}
The LP compiler~\cite{avis_polynomial_2019, avis_sparktope_2022} converts deterministic algorithms into LP models. If the algorithm runs in polynomial time and space, the resulting LP will be of polynomial size. 
The compiler can be viewed as a modeling language that defines subsets of the boolean hypercube---for certain 0/1 feasible points described through high-level programming languages---by generating linear constraints.
\begin{figure}[!bp] 
\begin{center}
\includegraphics [width=\columnwidth]{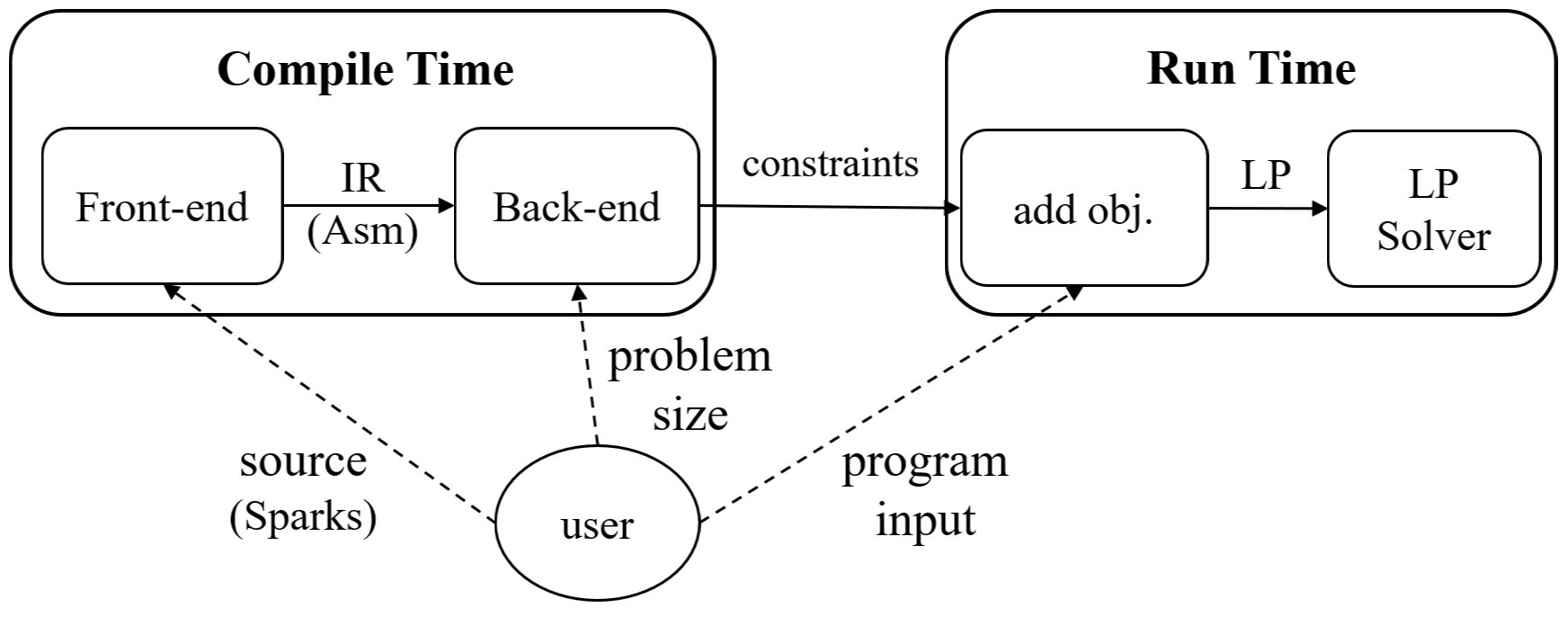}
\caption{The workflow of the LP compiler introduced in~\cite{avis_sparktope_2022}}
\label{fig:compiler-workflow}
\end{center}
\end{figure} 
For any polynomial-time computable \( f : \{0,1\}^p \rightarrow \{0,1\}^q \), there exists a polynomial-time computable \emph{imperative model} \( P = \{ z \mid Az \leq b \} \subseteq [0,1]^{p+q+r} \) such that for each input \( x \in \{0,1\}^p \), there exists a \emph{unique} vertex in the polytope corresponding to the triple \( (x, f(x), s) \in V(P) \cap \{0,1\}^{p+q+r}\), where \( s \) denotes auxiliary variables and \( V(P) \) is the set of all vertices of \( P \), which may contain non-integral vectors. This is equivalent to the \emph{$x$-0/1 property} discussed in~\cite{avis_polynomial_2019}. 

Figure~\ref{fig:compiler-workflow} illustrates the workflow of the LP compiler. At compile time, the user provides the problem size~$n$ and an upper bound on the number of execution steps for inputs of that size, denoted as $\TB(n)$. The generated LP formulation is general, as it can solve for any instance of size $n$. At run time, the user provides the specific input values, which are encoded into the LP via the objective function.
The imperative model represents source-level integer variables in binary as a set of $[0,1]$ constrained variables (i.e., not \enquote{Binary} in the \cplex{} terms).

To support the execution of intermediate-language statements---referred to as \Asm{}---at arbitrary time steps, the LP compiler versions all mutable \Asm{} variables with a time index $t$, and copies the relevant constraints for each statement across all $t \in 1,\dots,\TB(n)$. 
Integrality is assured through the propagation of fixed integer values, specified by initialization equalities and the objective function. The program state is tracked using boolean controller variables $S(l,t)$, which indicate that the implicit program counter (PC) is at line $l$ at time $t$. Execution is initialized at line 1 and time 1 via \(S(1,1)=1\) equality constraint.
Moreover, the set of constraints $\sum_{(l=1)}^{L} S(l,t)=1$, where $L$ is the total number of \Asm{} lines and $t \in 1,\dots,\TB(n)$, ensure that the PC is on exactly one line at each time step. 
Each program state has three types of constraints. The \emph{control flow} constraints determine which line becomes active in the next time step based on the current state. The \emph{carry-forward constraints} propagate the values of variables that remain unchanged. Lastly, the \emph{memory update} constraints apply the semantics of the active \Asm{} statement to update relevant variables. 

A well-known challenge in both the theoretical research and practical solution of IPs is that numerical solvers based on finite-precision floating-point arithmetic may return non-integral solutions, even when the actual optimal solution is integral~\cite{jarck_exact_2020}. 
Similarly, while the LP compiler is based on an imperative model that guarantees an integral optimal solution for $0/1$ input values, the LP formulations remain vulnerable to numerical precision errors introduced by current solvers. To validate correctness despite solver imprecision,~\cite{avis_sparktope_2022} explored two approaches: 
using exact arithmetic and fixing input variables to their 0/1 values. In the latter case, two out of three solvers successfully found the optimal 0/1 solution by determining the 0/1 values of the non-input variables, which are equivalent to the trace of the program execution for the given instance. These numerical issues are further discussed in Section~\ref{sec:results}. 

\section{Related Work}
\label{sec:relatedWork}
Methods for reducing LP size can be broadly categorized into modification and reformulation techniques.
\emph{Modification} techniques---commonly used by presolvers---aim to simplify and reduce LP size, improve numerical properties, detect infeasible constraints, and often lead to reducing solution times~\cite{swietanowski_modular_1995}. These techniques can be further classified based on whether they apply to the primal or dual of the LP~\cite{achterberg_presolve_2020, achterberg_mixed_2013}. 
Through their operations, presolvers enhance model sparsity---a desirable LP model characteristic---since the solution time often depends on the number of rows, columns, and nonzero entries in the coefficient matrix~\cite{andersen_presolving_1995}. While modification techniques can significantly reduce LP size in practice, their effect is generally limited to constant-factor reductions and does not change the asymptotic size of the LP. 

\emph{Reformulation} techniques generate polynomial-size LPs that are equivalent to the original exponential model, often by introducing a polynomial number of additional variables. One technique is Extended Formulations (EFs), which find equivalent polynomial-size LPs in higher dimensions such that the original exponential model can be recovered from a linear projection of the new one onto the original variable space~\cite{conforti_extended_2010, fiorini_smaller_2017, edmonds_paths_1965, fiorini_exponential_2015}.
Similarly, Compact Integer Programming (CIP) formulations can be used to avoid cutting plane methods for modeling constraints when solving exponential-size IPs. However, unlike EFs, CIPs do not necessarily project onto the natural polytope of the original problem~\cite{lancia_deriving_2016}. It is important to note that not all exponential polytopes have polynomial-size EFs. A notable example is the Edmonds' Matching Polytope for the Perfect Matching problem, which has been proven to have exponential extension complexity~\cite{rothvoss_matching_2017}. 

Alternative reformulation strategies based on separation oracles have also been introduced. 
Notably, Martin~\cite{martin_using_1991} reformulated the minimum spanning tree (MST) problem by encoding its polynomial-time separation oracle as a polynomial-size LP, resulting in the first compact formulation of MST. 
Similarly, a compact separation formulation allows a polynomial-size LP relaxation for an exponential-size IP, statically including the effect of all cuts that would otherwise be generated by the oracle. Its LP relaxation is as tight as the dynamic formulation that relies on an external oracle during optimization~\cite{carr_compact_2002}.
A recent reformulation approach by Avis et al.~\cite{avis_polynomial_2019} introduces a weaker notion of EF (WEF). This approach is implemented as an LP compiler in~\cite{avis_sparktope_2022} discussed in Section~\ref{sec:background}, which generates polynomial-size LPs directly from deterministic algorithms running in polynomial-time and space.
This work follows a line of research that traces back to the seminal work of Dobkin et al.~\cite{dobkin_linear_1979}, who proved that LP is $\PP$-complete by providing a log-space reduction from the Horn-SAT problem to an equivalent LP formulation. 
In a similar work, Valiant~\cite{valiant_reducibility_1982} showed that all problems in $\PP$ can be modeled with polynomial-size LPs by converting families of circuits into LPs.
However, as noted in Section~\ref{sec:introduction}, the LP compiler---which reformulates algorithms based on WEF---produces polynomial-size but extremely large LPs for problems in $\PP$, making them impractical for existing LP solvers. 
Although certain conventional compiler optimization techniques---such as \emph{dead code elimination}---can help reduce LP size~\cite{muchnick_advanced_1997, schneck_survey_1973}, many others have negligible impact or can even increase the size of the resulting LP. For instance, optimizations designed to increase data bandwidth~\cite{cardoso_compiling_2010} or apply \emph{loop unrolling} can significantly increase the LP size. 
Other techniques, such as code motion optimization---which reduces execution time by moving invariant computations to regions with lower dynamic frequency~\cite{schneck_survey_1973}---are ineffective in the LP compiler presented in \cite{avis_sparktope_2022}. 
This is because the compiler simulates code within the LP as if all statements execute with uniform frequency.

\section{Methodology}
\label{sec:methodology}
The proposed HSB method automatically decomposes code---guided by syntactic structure and type information---into a hierarchical, tree-like structure. In a single bottom-up pass, it annotates each block with time-distance metadata, expressed as constants or symbolic formulas, and determines its local $\TB(n)$ bounds. 
HSB modifies control flow by simulating synchronization barriers at the end of certain blocks, avoiding the overhead of extra time steps and code lines.
During constraint generation, it streams ETIs for each relevant block on demand from the annotated formulas, avoiding storage overhead. 
Likewise, during carry-forward constraint generation, it streams union of ETIs and transitions only for blocks that access the variables. By restricting constraint copies and variable versions to these intervals---often much smaller than the global $\TB(n)$---HSB drastically reduces LP size. The omitted constraints and variables are irrelevant in the HSB model due to the control-flow changes introduced by synchronization barriers.

\subsection{Synchronization Blocks (SBs)}
The compiler decomposes the source code into a hierarchical structure called the \emph{SB-tree}, consisting of uniquely labeled blocks called \emph{Synchronization Blocks} (SBs). 
Although the SB-tree has a tree structure, the relative ordering of its nodes reflects the execution semantics rather than the syntactic structure. 
The entire source code exists within the SB-tree leaves. 
SBs fall into several types, with the three main categories being loops, conditionals, and code blocks. These blocks are nested, though code blocks can also be flat. A flat code block consists of a sequence of one or more complete statements (e.g., an entire loop, but not a partial one).
Nested SBs of type loop or conditional include predefined sub-blocks for tasks such as bound checks and conditional evaluations.
\emph{Sequential} SBs are flat blocks resembling basic blocks---consisting of straight-line code with a single entry point and a single exit point---but allowing multiple exits through halts.

\subsection{Execution Time-Interval Generator (ETIG)}
In this section, we define the time-related constants and formulas that the compiler uses to annotate blocks in the SB-tree. These annotations enable efficient generation of ETIs. Throughout this section, we refer to the nested loop blocks as \emph{Loop Ancestors} (LAs) of their descendants.

\subsubsection{Time Distance to Parent ($\distParent$)}
Let $B$ denote the set of all SBs. For each $b \in B$, the compiler computes its local time bound $\tbLocal_b(n) \in \mathbb{N}$, based on compile-time parameters and code structure.
Let $\parent(b)$ denote the parent of block $b$, and let $\children(\parent(b))$ be the ordered list of its children. 
The set of $b$'s \emph{previous siblings} is defined as $\prevSiblings(b) = \{ b' \in \children(\parent(b)) \mid b' \prec b \}$, where $\prec$ denotes the compiler-defined ordering. This ordering considers both the position of blocks in the \Asm{} code and additional information related to execution timing. For instance, mutually exclusive blocks---such as \textbf{then} and \textbf{else} branches of an \textbf{if-else}---share the same execution time ranges and are not treated as siblings in the SB-tree.
The time offset of $b$ to its parent is defined as:
\begin{align*}
\distParent(b) = \left\{
\begin{aligned}
&\sum_{b' \in \prevSiblings(b)} \tbLocal_{b'}(n) && \text{if } \prevSiblings(b) \ne \emptyset \\
&0 && \text{if } \prevSiblings(b) = \emptyset
\end{aligned}
\right.
\end{align*}
\subsubsection{Penultimate Time Distance ($\distPenultimateToSA$)}
For each $\vLA \in B$ which is an LA block, let $\distPenultimateToSA(\vLA, i) \in \mathbb{N}$ be the \emph{penultimate time distance} of the $i$th iteration of $\vLA$. This value represents the time offset from the start of $\vLA$ to the end of its $(i - 1)$th iteration. Accordingly, the $i$th iteration of $\vLA$ begins at offset $\distPenultimateToSA(\vLA, i) + 1$. The compiler annotates each LA node in the SB-tree with its $\distPenultimateToSA$ formula. For instance, if $\vLA$ is a \textbf{for} loop, the compiler stores the formula: 
\[
\distPenultimateToSA(\vLA, i) = \tbLocal_{\BONE}(n) + (i-1) (\tbLocal_{\BTWO}(n) + \tbLocal_{\BTHREE}(n))
\]

Here, $\BONE$ and $\BTHREE$ are predefined SBs specific to \textbf{for} loops (following the design in~\cite{avis_sparktope_2022}), and $\BTWO$ is the nested block representing the body of the loop. 
The values $\tbLocal_{\BONE}(n)$, $\tbLocal_{\BTWO}(n)$, and $\tbLocal_{\BTHREE}(n)$ are compile-time constants. Only $i$ is a symbolic variable, substituted with actual index values during LP constraint generation.
\subsubsection{Cumulative Time Distance ($\distCumulativeToTarget$)} 
The \emph{cumulative time distance} of block $b$, denoted $\distCumulativeToTarget(b)$, is a compile-time constant representing the time offset of $b$ from the root of the SB-tree, assuming the first iteration of all $b$'s LAs. 
Let \emph{non-root ancestors}, denoted $\nonRootAncestors(b)$, refer to the set of all ancestors of $b$ excluding the root.
Then the $\distCumulativeToTarget$ of $b$ is computed as:
\[
\distCumulativeToTarget(b) = \distParent(b) + \sum_{a \in \nonRootAncestors(b)} \distParent(a)
\]
\subsubsection{Execution Time Interval (ETI)}
\label{sec:ExecutionTimeIntervalETI}
Let $b \in B$ be a target block nested within $k$ LAs, denoted $\vLA_1, \dots, \vLA_k$, where each $\vLA_i \in B$ and has a maximum iteration bound $\vMi_i$. 
A specific execution of the nested loops is specified by a tuple of loop indices $J = (j_1, \dots, j_k)$, where $j_i \in [1,\vMi_i]$. The start and end times of the ETI of $b$, corresponding to the iteration $J$, are defined as: 
\begin{align*}
\ETIstart(b, J) &= \sum_{i=1}^{k} \distPenultimateToSA(\vLA_i, j_i) + \distCumulativeToTarget(b) \\
\ETIend(b, J) &= \ETIstart(b, J) + \tbLocal_b(n) \\
\end{align*}
Therefore, the ETI of $b$ for the index tuple $J$ is:
\[
\ETI(b, J) = [ \ETIstart(b, J),\ \ETIend(b, J)]
\]
\subsubsection{Index Generation via Loop Fusion}
Target blocks may be nested within different numbers of LAs. To handle this variability, ETIG models the index generation as a \emph{mixed-base counting} problem. In the model, each digit corresponds to an LA and has a base equal to its maximum iteration bound $\vMi$. Consequently, index tuples are treated as digits of a mixed-base integer of width $k$, one-padded on the left.
Consider a block $b \in B$ nested within $k$ LAs. Let $\vMaxIterTuple_b = (\vMi_1, \dots, \vMi_k)$ be the tuple of maximum iteration bounds of LAs. The $\ETIG(b, \vMaxIterTuple_b)$ does not return all $\prod_{i=1}^{k} \vMaxIterTuple_{b,i}$ ETIs at once, but instead yields each $\ETI(b, J)$ on demand, where $J = (j_1, \dots, j_k)$ and $j_i \in [1,\ \vMaxIterTuple_{b,i}]$ for each $i \in [1, k]$. 

\subsection{Synchronization Barrier Constraints }
\label{sec:syncBarrierConstraints}
Control flow exiting certain SBs---such as non-sequential flat code blocks whose execution time may vary depending on input data---is directed to an \emph{idle line} at the end of the block, where it remains until its current time matches the end of the active ETI. To avoid an extra time step for idling when the flow has already reached the end of ETI, the compiler inserts predefined \emph{flow blocks} at the end of certain block types, such as \textbf{else} branches. These additions allow the compiler to locally detect, when generating control flow constraints for blocks within nested structures, whether to generate constraints for idling or to model direct transitions to the next block. 

\subsection{Efficient Constraint Generation }
To reduce the number of LP columns, the compiler finds the union of ETIs for flat SBs that access an \Asm{} variable. 
This union is generated by the Union Execution Time Interval Generator (UETIG), described in the next section. 
Variable versions are then created only for the resulting time intervals. While further reduction is possible---potentially generating variable versions only for time steps within the ETI of the blocks that modify the \Asm{} variable---this paper implemented a simpler strategy. Specifically, a full carry-forward is performed, which propagates the variable values step-by-step across each ETI, while also adding an extra version before each interval. This method avoids using UETIG when generating other constraints, such as those related to conditional control flow and memory updates. 
The compiler further reduces the number of LP rows by eliminating control-flow and memory-update constraints at time steps that do not fall within a block's ETIs. This reduction is particularly significant for sequential blocks, where the compiler emits a single set of constraints per ETI, rather than one for every time step within the interval. 
To reduce constraints introduced by multiple return statements---which may include boolean operations---the compiler inserts an additional halt line that serves as a unified control-flow target.  
This redirection enables LP size reduction for blocks containing the return statements, as constraints are generated only for their ETIs. 
However, the halt line preserves the imperative model's properties described in Section~\ref{sec:background}.

\begin{figure}[!bp] 
\begin{center}
\includegraphics [width=\columnwidth]{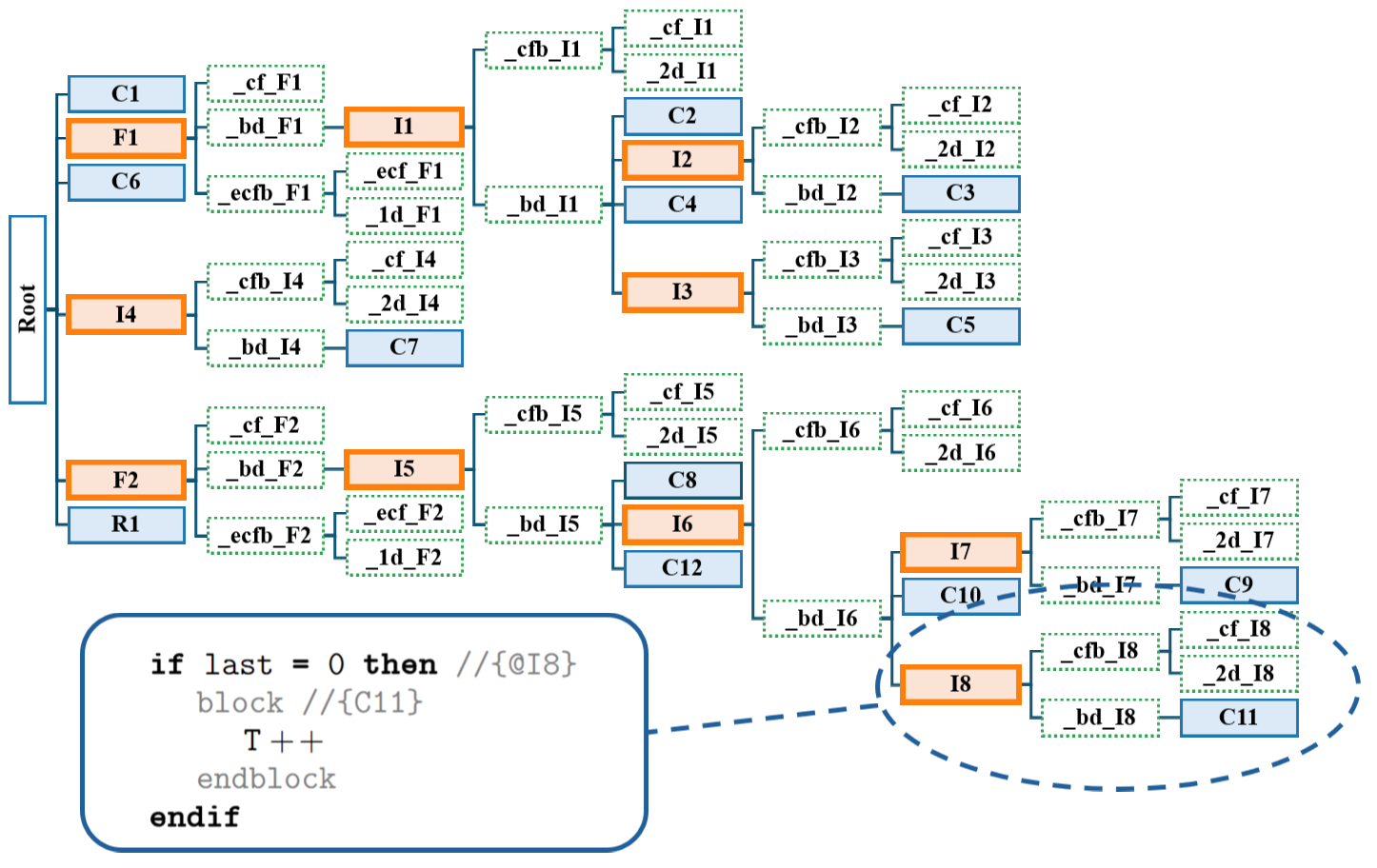}
\caption{SB-tree generated for the makespan algorithm from~\cite{avis_sparktope_2022}, shown alongside the automatically annotated \Sparks{} code corresponding to the sub-tree rooted at block $\iEight$. Node styles indicated SB types: green dotted nodes represent compiler-added predefined sub-blocks, bold orange nodes indicate nested loop and conditional blocks SBs that end with synchronization barriers; and solid border blue nodes denote flat sequential code blocks.}
\label{fig:sbtree-makespan}
\end{center}
\end{figure}

Figure~\ref{fig:sbtree-makespan} 
illustrates the SB-tree automatically generated by the compiler for the makespan algorithm from~\cite{avis_sparktope_2022}, along with an annotated snippet of the corresponding \Sparks{} code for the sub-tree rooted at the block labeled $\iEight$. 
In the figure, green dotted nodes represent compiler-added predefined sub-blocks, plain solid blue nodes represent flat sequential blocks, and the bold orange nodes are nested loops and conditional blocks containing synchronization barriers.
The user provides the maximum number of iterations, $\vMi$, only for the two loop blocks labeled $\fOne$ and $\fTwo$; the compiler then automatically computes the local $\TB(n)$ of each block, with the $\TB(n)$ of the root block defining the global $\TB(n)$ of the code. 

\subsection{Union Execution Time Interval Generator (UETIG)}
\label{sec:UnionExecutionTimeIntervalCaclulatorUETIC}
The UETIG incrementally yields ETIs during which any block in a given set $\{ b_1, \dots, b_u \} \subseteq B$, each of which accesses a particular \Asm{} variable $v$, may execute. It performs a time-based incremental search over the range $1$ to $\TB(n)$, returning ETIs that contain the current time $t$, and advancing $t$ by the duration of each yielded ETI.
For each $b_i$, let $\vMaxIterTuple_{b,i}$ be the tuple of maximum iteration bounds of its LAs, and let $\ETIG(b_i, \vMaxIterTuple_{b,i})$ be its corresponding ETI Generator.
The compiler builds a list of ETIGs of blocks, denoted $\mathcal{G} = [\ETIG(b_1, \vMaxIterTuple_{b,1}), \dots, \ETIG(b_u, \vMaxIterTuple_{b,u})]$. UETIG starts at time $t=1$, and for each generator $\ETIG(b_i, \vMaxIterTuple_{b,i}) \in \mathcal{G}$, it updates their ETI, denoted $\ETI(b_i, J_i)$, until either $t \in \ETI(b_i, J_i)$, $\ETIstart(b_i, J_i) > t$, or no ETIs remain in $\ETIG(b_i, \vMaxIterTuple_{b,i})$.
If there exists a flat block $b_r$ such that $t \in \ETI(b_r, J_r)$, UETIG yields $\ETI(b_r, J_r)$ and updates the time to $t = \ETIend(b_r, J_r)$. It then resumes by updating the ETIs based on the new time $t$ and continues yielding the next interval. The process terminates when $t > \TB(n)$ or when all ETIGs are exhausted. 

\section{Results}
\label{sec:results}
This section compares the LP size and solver performance of LPs generated by our proposed HSB method and the unoptimized LP compiler (UO)---the compiler from~\cite{david_avis_sparktope_2020}, on which our implementation is based. We evaluate both approaches on two benchmark problems, makespan and minimum spanning tree (MST).
All experiments were performed on a system equipped with Intel(R) Xeon(R) Gold 6248 CPUs running at 2.50 GHz.
\subsection{Linear Program Size Reduction}
\begin{figure}[bp] 
\centering
\resizebox{\columnwidth}{!}{%
\includegraphics [width=\columnwidth]{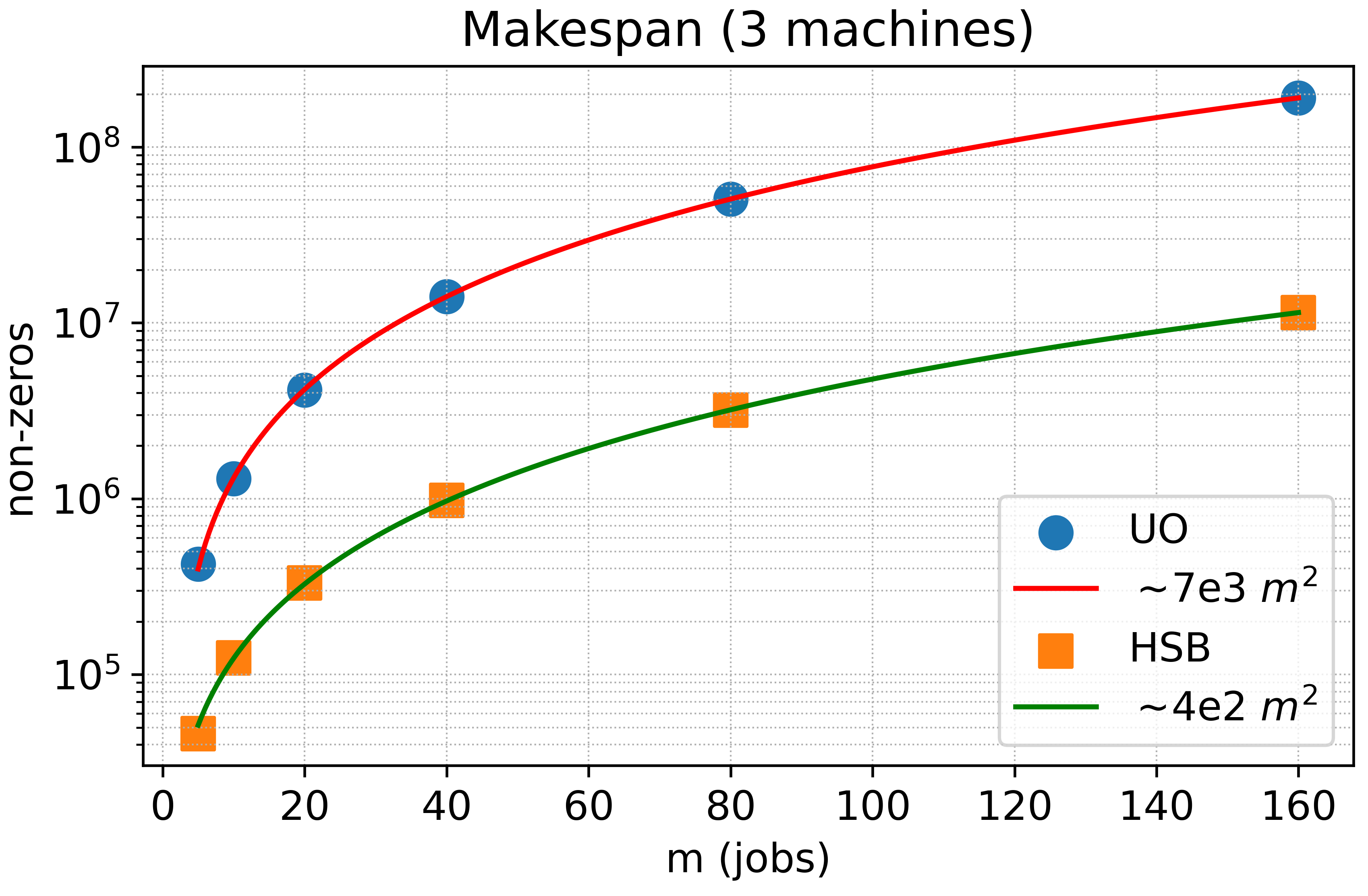} 
}
\caption{Number of non-zeros in LPs for the makespan problem with $n=3$ machines, comparing the unoptimized compiler (UO) and the Hierarchical Synchronization Barriers (HSB) method.}
\label{fig:funcfit-nonzeros-makespan}
\end{figure}

Tables~\ref{tab:lp-sizes-ms} and~\ref{tab:lp-sizes-mst} report the LP sizes for the makespan and MST problems, respectively. Across all tested input sizes, the proposed HSB method generates substantially smaller LPs than UO. The reductions are especially significant for the MST problem, which operates over weighted graphs and includes integer array operations. 
For the largest input size of the makespan problem (Table~\ref{tab:lp-sizes-ms}), HSB achieves reductions of approximately $94\%$ in the number of non-zeros and constraints, $93\%$ in LP file size, and $53\%$ in the number of variables.
As shown in Figure~\ref{fig:funcfit-nonzeros-makespan}, where we fit a quadratic function to the number of non-zero elements, HSB achieves a substantial 17.5 times reduction in the leading coefficient of the non-zero count, without altering the asymptotic growth rate. This behavior is expected and aligns with optimization techniques in compilers and LP presolvers, which typically achieve practical improvements without changing the asymptotic runtime or asymptotic LP sizes (as discussed in Section~\ref{sec:introduction}). 
\begin{table}[bp] 
\centering
\caption{LP sizes for the makespan problem with $m$ jobs on $n=3$ machines, comparing outputs from the unoptimized compiler (UO) and the Hierarchical Synchronization Barriers (HSB) method. Rows denote constraints and columns denote variables of the LP model.}
\label{tab:lp-sizes-ms}
\resizebox{\columnwidth}{!}{%
\begin{tabular}{c c c c c c c}
\toprule
\makecell{$m$} & 
\makecell{opt.} & 
\makecell{$\TB(n)$} &
\makecell{rows\\$(\times 1000)$} & 
\makecell{columns\\$(\times 1000)$} & 
\makecell{non-zeros\\$(\times 1000)$} & 
\makecell{file\\(MB)} \\
\midrule
\multirow{2}{*}{5}
& UO & 201  & 129 & 27 & 426 & 8 \\ 
 & HSB & \textbf{194} & \textbf{8} & \textbf{16} & \textbf{46}  & \textbf{1} \\ 
\midrule
\multirow{2}{*}{10}
  & UO & \textbf{321} & 380 & 58 & 1,304 & 23  \\ 
  & HSB & 374 & \textbf{25} & \textbf{39} & \textbf{124} & \textbf{3}  \\
\midrule
\multirow{2}{*}{20}
  & UO & \textbf{631} & 1,170 & 148 & 4,160 & 75  \\ 
 & HSB & 734 & \textbf{74} & \textbf{92} & \textbf{331}  & \textbf{7}  \\
\midrule
\multirow{2}{*}{40}
  & UO & \textbf{1,251} & 3,845 & 411 & 14,099 & 255  \\ 
 & HSB & 1,454 & \textbf{236} & \textbf{232} & \textbf{976} & \textbf{21} \\
\midrule
\multirow{2}{*}{80}
 & UO & \textbf{2,491} & 13,483 & 1,250 & 50,634 & 936  \\ 
 & HSB & 2,894 & \textbf{813} & \textbf{636} & \textbf{3,196} & \textbf{69}  \\
\midrule
\multirow{2}{*}{160}
  & UO & \textbf{4,971} & 49,869 & 4,150 & 190,407 & 3,595 \\ 
 & HSB & 5,774 & \textbf{2,999} & \textbf{1,923} & \textbf{11,445} & \textbf{249} \\ 
\bottomrule
\end{tabular}
}
\end{table}

\begin{figure}[!bp] 
\centering
\resizebox{\columnwidth}{!}{%
\includegraphics [width=\columnwidth]{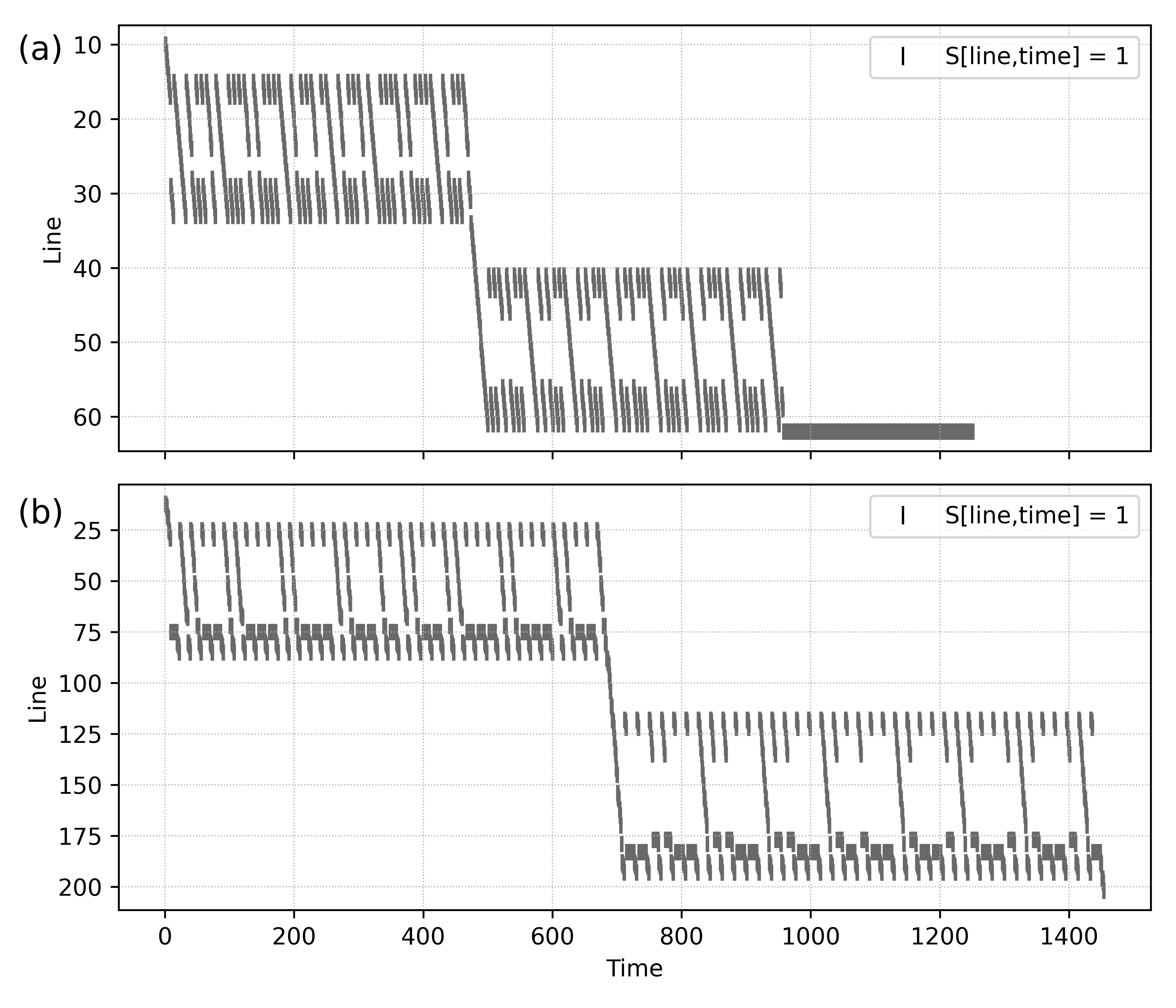} 
}
\caption{Execution traces of the makespan algorithm with 3 machines and 40 jobs. Subfigures show: (a) Unoptimized compiler (UO). (b) Hierarchical Synchronization Barriers (HSB).}
\label{fig:trace-ms40}
\end{figure}

\begin{table}[tp] 
\centering
\caption{LP sizes for the minimum spanning tree (MST) problem with input size $n$, generated using the unoptimized LP compiler (UO) and with the Hierarchical Synchronization Barriers (HSB) optimization. Rows denote constraints and columns denote variables of the LP model.} 
\label{tab:lp-sizes-mst}
\resizebox{\columnwidth}{!}{%
\begin{tabular}{c c c c c c c}
\toprule
\makecell{$n$} & 
\makecell{$\TB(n)$} &
\makecell{opt.} & 
\makecell{rows\\$(\times 1000)$} & 
\makecell{columns\\$(\times 1000)$} & 
\makecell{non-zeros\\$(\times 1000)$} & 
\makecell{file\\(MB)} \\
\midrule
\multirow{2}{*}{3}
 & \multirow{2}{*}{573} & UO & 1,277 & 182 & 4,342 &  84 \\
 &  &HSB & \textbf{55}    & \textbf{73}    & \textbf{279}   & \textbf{6}  \\
\midrule
\multirow{2}{*}{5}
  & \multirow{2}{*}{1,753} & UO & 8,059 & 911 & 28,726 &  564 \\
  &  & HSB & \textbf{263}   & \textbf{249}   & \textbf{1,185} & \textbf{32}  \\
\midrule
\multirow{2}{*}{8}
  & \multirow{2}{*}{5,323} & UO & 39,569 & 4,221 & 146,047 &  2,926 \\
 &  & HSB & \textbf{978}   & \textbf{780}   & \textbf{4,266} & \textbf{139}  \\
\midrule
\multirow{2}{*}{12}
  & \multirow{2}{*}{14,703} & UO & 247,862 & 10,337 & 941,313 & 19,395  \\
 &  & HSB & \textbf{4,307} & \textbf{2,418} & \textbf{17,685} & \textbf{771}  \\
\bottomrule
\end{tabular}
}
\end{table}
HSB's reductions become increasingly pronounced as algorithm complexity increases. As shown in Table~\ref{tab:lp-sizes-mst}, the MST algorithm benefits even more from HSB. For the largest input size, HSB reduces the number of non-zeros and constraints by approximately $98\%$, the file size by $96\%$, and the number of variables by $77\%$.
These larger reductions, especially in the number of variables, result from the Prim's algorithm's nested loop structure and the higher concentration of \Asm{} variables within fewer blocks. This structure enhances HSB's effectiveness, as it generates constraint copies and variable versions for each statement only across the time steps included in the ETIs of their enclosing blocks.
Figure~\ref{fig:trace-ms40} illustrates execution traces of the makespan problem for the unoptimized compiler (UO) and the HSB optimization. 
Each trace shows the time steps at which the $S$ controllers are set to one in the optimal solution, indicating which line of code executes at each time step. In Figure~\ref{fig:trace-ms40}(a), the UO trace completes execution around $950$ and remains at the return line for the remainder of its $\TB(n)$ time bound. In contrast, Figure~\ref{fig:trace-ms40}(b) shows that HSB introduces idle time at the end of certain blocks through synchronization barriers, producing a more elongated trace. 
When HSB and UO use the same upper bound $\TB(n)$, the differences in their execution traces primarily reflect a redistribution of UO's idle time---originally concentrated on the return line---across multiple HSB blocks. However, as discussed in the next section, in cases where UO achieves a smaller $\TB(n)$ that HSB cannot match, the trace differences are not merely due to idle time redistribution. Instead, they also result from additional idle time introduced by HSB's synchronization constraints, idle time that is absent from the UO trace.

\subsection{Custom $\TB(n)$ Reductions}
Although the $\TB(n)$ values of HSB and UO are similar for some algorithms, such as Prim's algorithm using an adjacency matrix (Table~\ref{tab:lp-sizes-mst}), custom $\TB(n)$ reductions based on user insight are possible for some algorithms, such as makespan algorithm presented in~\cite{avis_sparktope_2022}. This problem-specific $\TB(n)$ reduction can result in smaller LPs for UO.
For example, the makespan algorithm consists of two unnested loops, each iterating $m$ times and containing conditional statements. The user has the knowledge that the conditional bodies across both loops execute a total of $m$ times. Specifically, if the conditional body in the first loop executes $u$ times, then the one in the second loop executes $m-u$ times, for some $0 \leq u \leq m$.
UO can leverage such custom $\TB(n)$ reductions because its generated LPs allow statements to execute at any time step within the overall $\TB(n)$. 
In contrast, HSB cannot apply custom $\TB(n)$ reductions, as it structures the code hierarchically and synchronizes between blocks based on their local $\TB(n)$ values, effectively modeling the worst-case execution time of each block. 
As shown in Table~\ref{tab:lp-sizes-ms}, UO achieves consistently smaller $\TB(n)$ values than HSB, with the only exception occurring at $m = 5$. This inconsistency is likely due to an error in~\cite{avis_sparktope_2022}, where the correct value should be $166$ based on the formula provided in that paper---still smaller than the value produced by HSB.

HSB's larger $\TB(n)$ values result from the fact that conditional bodies in each loop must either execute or idle during all $m$ iterations of their loop, leading to $2m$ iterations in total compared to only $m$ in UO.
Despite this, HSB still generates substantially smaller LPs in the makespan case. This is because the variable versions and constraint copies associated with each statement are confined to the ETIs of the statement's enclosing block. The ETIs, determined based on the local $\TB(n)$ of blocks, typically occupy a much smaller subset of the overall $\TB(n)$ used by UO.
However, HSB becomes ineffective when custom $\TB(n)$ reductions are significant. For example, in the maximum matching problem from~\cite{avis_sparktope_2022}, the user has the knowledge that the block responsible for the shrinking operation in the blossom algorithm executes only once per outermost loop iteration.
This block is deeply nested and has a very large local $\TB(n)$, resulting in a significant reduction of the overall $\TB(n)$. Because HSB relies on worst-case execution times for each block, it cannot exploit this reduction. The resulting gap between HSB's $\TB(n)$ and the reduced $\TB(n)$ used by UO becomes so large that HSB generates a larger LP than UO. 

That said, HSB could be extended to support certain custom $\TB(n)$ reductions---especially when the reduction patterns are expressible in terms of compile-time parameters, as seen in the makespan and maximum matching algorithms. 
Moreover, when further nesting of an HSB block is undesirable---such as preventing application of a problem-specific reduction in $\TB(n)$---the user can manually flatten the block to enable local $\TB(n)$ reduction at that level. In such cases, the effectiveness of HSB diminishes in proportion to the share of the total $\TB(n)$ contributed by the flattened block. The larger that share, the smaller the overall benefit HSB provides for the program.
Nonetheless, certain compiler optimizations can enhance the effectiveness of HSB even in such scenarios. For instance, in the maximum matching problem, HSB generates significantly smaller LPs when a simple code motion optimization (Section~\ref{sec:relatedWork}) moves the shrinking operation to the outermost loop of the source code. This improves HSB's performance because it generates constraints proportional to the number of execution steps within a block's ETIs---which is considerably smaller when the block is no longer nested within multiple loops.

\subsection{Solver Performance}
\label{sec:solverPerformance}
We evaluate solver performance using two commercial solvers---\cplex{} 22.1.1.0 and \gurobi{} 12.0.1---and one non-commercial solver, \scip{} 9.2.1. 
All solvers run with default settings, subject to the computational resource constraints detailed in the corresponding figure captions. For each configuration, we report the median elapsed time, presolve time, and memory usage across multiple runs. We omit CPU time, as it closely matches elapsed time. We also omit interquartile range (IQR) and standard deviation, as their values are negligible across all results, with the exception of a single case discussed below.

\begin{figure}[!htbp] 
\centering
\resizebox{\columnwidth}{!}{%
\includegraphics [width=\columnwidth]{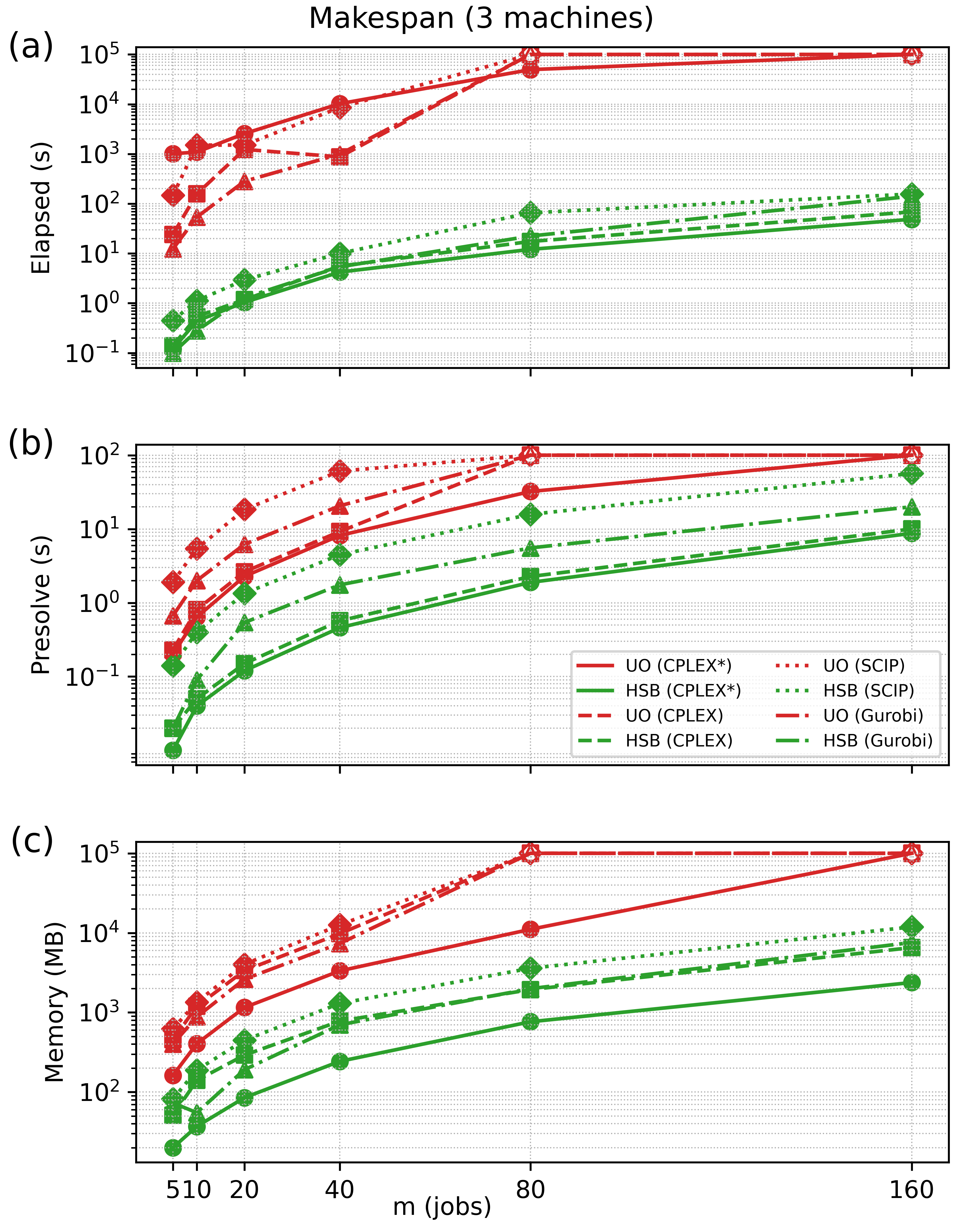} 
}
\caption{Solver performance on the makespan problem with 3 machines, where inputs are encoded via the objective function. LPs are generated using the unoptimized compiler (UO) and with the Hierarchical Synchronization Barriers (HSB) optimization. A superscript $^*$ indicates the use of the dual simplex algorithm. 
Each data point represents the median of 10 runs, each subject to a 23-hour time limit, 32~GB memory, and 7 CPUs.
Unfilled markers at the top of each plot indicate solver failure due to exceeding the memory limit. 
Subfigures show: (a) elapsed time, (b) presolve time, and (c) memory usage.
}
\label{fig:msSolverPerformance}
\end{figure}

\begin{figure}[!htbp]
\centering
\resizebox{\columnwidth}{!}{%
\includegraphics [width=\columnwidth]{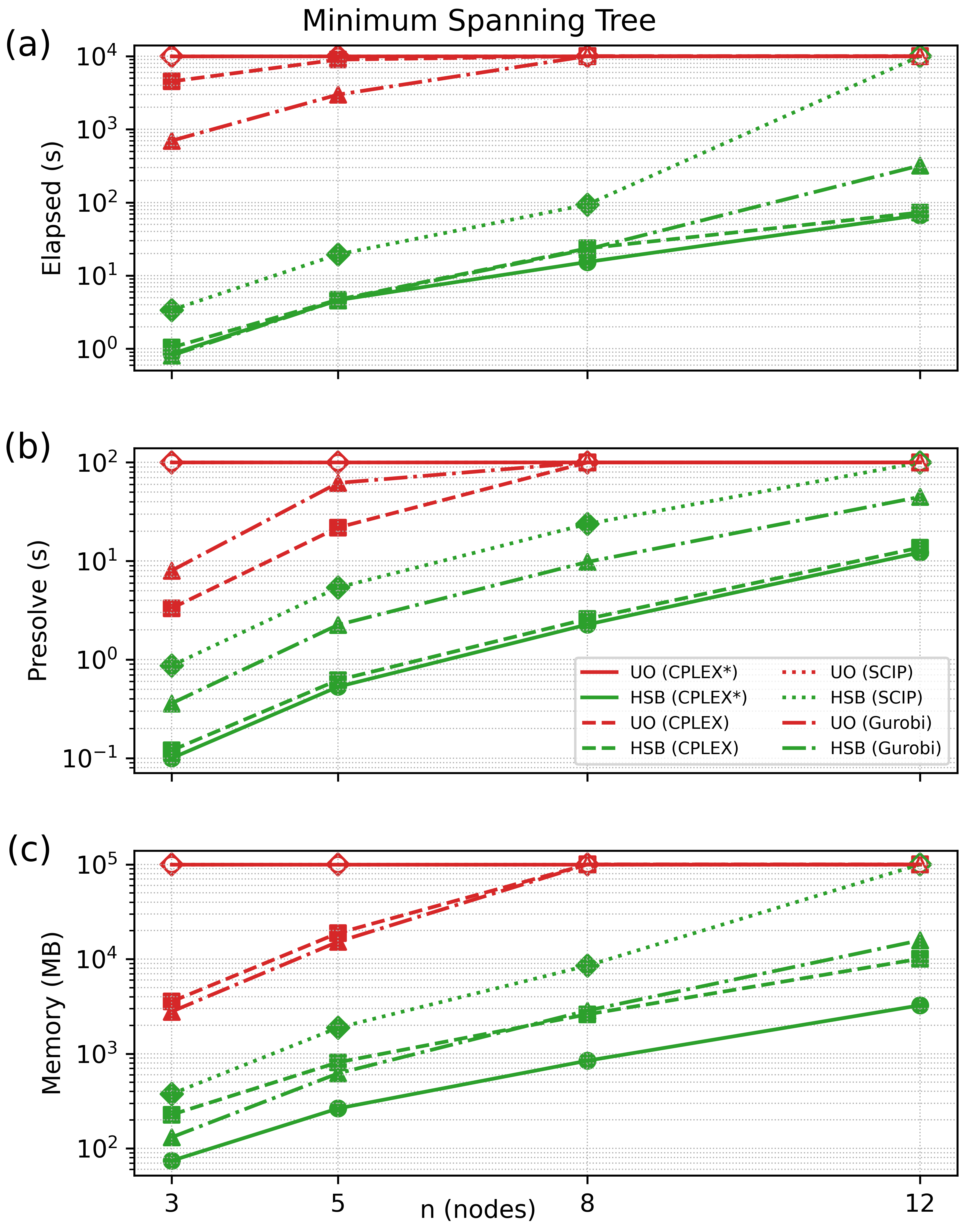} 
}
\caption{Solver performance for the minimum spanning tree (MST) problem, where inputs are encoded via the objective function. LPs are generated using the unoptimized compiler (UO) and with the Hierarchical Synchronization Barriers (HSB) optimization.
A superscript $^*$ indicates the use of the dual simplex algorithm. 
Each data point represents the median of 10 runs, each subject to a 23-hour time limit, 32~GB memory, and 7 CPUs.
Unfilled markers at the top of each plot indicate solver failure due to exceeding the memory or time limit.
Subfigures show: (a) elapsed time, (b) presolve time, and (c) memory usage.
}
\label{fig:mstSolverPerformance}
\end{figure}

Figures~\ref{fig:msSolverPerformance} and~\ref{fig:mstSolverPerformance} 
show solver performance---measured by elapsed time, presolve time, and memory usage---for the makespan and MST problems, respectively. 

Across all solvers and input sizes, HSB consistently outperforms UO by generating not only smaller LPs but also instances that are significantly easier to solve.
Solver performance is generally stable across runs for both UO and HSB, resulting in low IQRs across all runs. The only notable exception is UO on the makespan problem with $m=5$, where the dual simplex variant of \cplex{}, denoted as \cplex$^*$, shows an IQR equal to $32.08\%$ of the median elapsed time, indicating variability in solver performance.
Among the solvers, \cplex{} and \gurobi{} show comparable performance, with \cplex{} slightly outperforming \gurobi{} on larger LPs. As input size increases, solvers often reach the 23-hour time limit or $32$~GB memory cap for UO LPs. In contrast, HSB LPs remain solvable---typically within seconds or minutes, even for the largest input size.
The dual simplex variant, \cplex$^*$, performs poorly on UO LPs for the makespan problem and fails to solve any UO instance of the MST problem.
However, it performs exceptionally well on HSB LPs, in some cases outperforming all other solvers. 
For instance, in Figure~\ref{fig:msSolverPerformance}(a), \cplex$^*$ solves the HSB instance with $m=5$ more than $8,400$ times faster than the corresponding UO LP---a $99.9\%$ reduction in elapsed time.
Furthermore, comparing UO at $m=40$ with HSB at $m=160$---two instances with nearly identical file sizes ($255$ MB vs. $249$ MB)---all solvers solve the HSB instance between $7$ and over $210$ times faster. This consistent performance gap suggests that UO LPs may have structural disadvantages that do not align well with the algorithms and strategies employed by these solvers.
Additionally, solvers running with default settings---such as \cplex{}, \gurobi{}, and \scip{}---typically execute multiple algorithmic strategies in parallel to identify the most effective method for solving a given LP. This approach is specifically effective for large or complex instances but has the disadvantage of increased memory usage. In contrast, solvers configured to use a specific non-parallel algorithm (e.g., \cplex$^*$ with the dual simplex method) consume significantly less memory, and may succeed in solving instances where other solvers fail due to memory exhaustion. For example, in the makespan problem at $m=80$, only \cplex$^*$ can solve the UO instance. 
While HSB LPs remain solvable across all tested input sizes, UO LPs can only be solved up to $m=40$. At this size, \gurobi{} and \cplex{} solve HSB LPs $184$ and $156$ times faster, respectively, compared to the UO LPs.

For the MST problem, Figure~\ref{fig:mstSolverPerformance} presents solver performance across input sizes. \scip{} and \cplex$^*$ fail to solve any UO instances, while \gurobi{} and \cplex{} are able to solve UO LPs only up to $n=5$ before reaching the memory limit. In contrast, HSB instances remain solvable across all tested input sizes for all solvers, with the only exception of \scip{}, which reaches the memory limit at $n=12$. 
The performance gap is particularly noticeable at $n=5$, where \gurobi{} and \cplex{} solve the HSB instance approximately $645$ and $1912$ times faster, respectively, than the corresponding UO LP.
Since the inputs in Figures~\ref{fig:msSolverPerformance} and~\ref{fig:mstSolverPerformance} are encoded solely through the objective function---rather than being forced to $0/1$ values---most solvers returned fractional solutions. This behavior is consistent with previous observations in~\cite{avis_sparktope_2022}, as discussed in Section~\ref{sec:background}. Notably, the makespan instance with $m=5$ is now solvable by \scip{}, a non-commercial solver designed with a focus on solving Mixed-Integer Programs (MIPs). 
Moreover, when the input variables are fixed to their $0/1$ values, the commercial solvers \cplex{} and \gurobi{} successfully find optimal solutions, except in cases where they reach the applied memory or time limits. In contrast, \scip{} returns fractional solutions for the makespan problem at $m = 80$ and $m = 160$, and for the MST problem at $n=8$ when using the HSB optimization. This behavior does not indicate a flaw in HSB relative to UO, as \scip{} runs out of memory for the UO version on much smaller instances of both problems. Overall, compared to \cplex{} and \gurobi{}, \scip{} appears to lack certain presolving techniques that are effective for this type of LPs.
Promisingly, \gurobi{}, which in its earlier versions was reported to return non-optimal fractional solutions~\cite{avis_sparktope_2022}, now consistently finds optimal solutions. This improvement likely reflects advances in numerical precision and presolving techniques, enabling the solver to more effectively handle such LPs. 
From the perspective of our long-term goal of embedding these LPs within IP formulations, one possible approach to mitigate solver limitations is to declare the input variables as integers when using MIP solvers.

\section{Conclusion}
\label{sec:conclusion}
Linear Programming (LP) continues to be widely applied in industry for solving real-world problems having well-structured linear constraints, often outperforming or complementing data-driven approaches such as machine learning. It has played a critical role in the life cycle of many products, from design and manufacturing to logistics and transportation.
In this paper, we propose efficient methods for compiling algorithmic descriptions into significantly smaller LPs. This interface allows practitioners to model LPs using intuitive high-level programming languages, avoiding the need to manually specify LP constraints from scratch in Algebraic Modeling Languages (AMLs). 
Additionally, it enables machines to systematically generate Compact Integer Programming (CIP) formulations for problems whose Integer Programs (IPs) are exponential in size, provided they have polynomial-time separation oracles. In such cases, the oracle logic can be embedded directly into the polynomial portion of the IP model. 
We also aimed to generate significantly smaller LP formulations for problems that only have natural LP formulations with exponential extension complexity. On benchmark problems such as makespan and minimum spanning tree, our method generated LPs that are smaller by several orders of magnitude and consistently easier to solve across both commercial and non-commercial solvers. 
Unlike traditional compiler optimizations and LP presolve techniques, our approach exploits both the knowledge of the source code structure and the design of the polytope. It decomposes the code into a hierarchy of smaller, statically analyzable phases. Through interval abstraction and synchronized transitions, we reduce compile-time uncertainty over program execution patterns, eliminating redundant LP constraints and variables corresponding to unreachable or irrelevant states.
A promising direction for future work is to study the impact of these integrated CIPs on existing MIP solver performances in the absence of external oracles---a particularly intriguing question given the highly variable time and space usage of solvers on different models of the same size.

\bibliographystyle{plain}
\bibliography{hsb.bib}

\end{document}